\documentclass[twocolumn]{article}

\usepackage{siunitx}
\usepackage{graphicx}
\usepackage{authblk}

\usepackage[backend=biber]{biblatex}

\title{Microstructural Characterization of $Nb_3Sn$ Thin Films Using FIB Tomography}

\author[1]{Eric Viklund\thanks{Now working at KEK.}\thanks{Corresponding author: eric.viklund@kek.jp}}
\author[1]{David N. Seidman}
\author[1]{Sam Posen} 

\affil[1]{Fermi National Accelerator Laboratory}

\begin{document}

\maketitle

\begin{abstract}
The accelerating gradient of $Nb_3Sn$ superconducting radiofrequency (SRF) cavities is currently limited, and the underlying cause remains an open question in the field. One leading hypothesis attributes this limitation to the presence of tin-deficient regions within the $Nb_3Sn$ coating, which can suppress the superheating field. Due to the relatively large coherence length of $Nb_3Sn$, defects near the surface may significantly interact with the RF field. However, these subsurface defects have proven difficult to characterize. This research aims to investigate the structure and distribution of subsurface Sn deficient regions to better understand their influence on cavity performance. We employ focused ion beam (FIB) tomography to analyze the subsurface microstructure of $Nb_3Sn$ thin films. This technique enables three-dimensional reconstruction of both the tin distribution and the grain structure within the film. By correlating Sn content with grain structure, we find that Sn deficient regions are more prevalent that previously thought. However, the Sn deficient regions are consistently located below the surface of the film where RF fields are strongly attenuated by supercurrent screening and are likely not a limiting factor for cavity performance.
\end{abstract}

\section{Introduction}

$Nb_3Sn$ is a promising material for superconducting radiofrequency (SRF) cavities due to its high critical superconducting transition temperature ($T_c$) and high superheating magnetic field ($H_{sh}$) compared to Nb. $Nb_3Sn$ cavities can theoretically achieve a much higher accelerating gradient and lower surface resistance than Nb ones. However, in practice $Nb_3Sn$ cavities are highly sensitive to surface defects and struggle to reach the accelerating gradients achieved by Nb cavities. These defects stem from the manufacturing process of $Nb_3Sn$ cavities. $Nb_3Sn$ is a brittle material and needs to be deposited as a thin film on a structural substrate, typically a Nb cavity, using a coating method such as Sn vapor diffusion. This requirement leads inevitably to more defects compared to manufacturing from bulk materials. 

One type of defect known to occur in Sn vapor diffusion coated films is Sn-deficient regions.\cite{beckerAnalysisNb3SnSurface2015,leeAtomicscaleAnalysesNb3Sn2018,trenikhinaPerformancedefiningPropertiesNb3Sn2017} Becker et al first revealed the presence of Sn deficient regions using 2D cross-section imaging. However, several important questions are raised by these studies. Are the regions so close to the surface that they would be exposed to significant RF currents? Are they indeed located far from grain boundaries, indicating a mechanism for their formation, or are the 2D slices giving an incomplete picture? These important questions remain unanswered since the initial studies. Depending on how the regions are distributed in the film, they could potentially be critical defects, or they could be relatively benign. 

The effect of Sn deficiency is to lower the $T_c$ of the material which has an adverse effect on the cavity performance\cite{sitaramanEffectDensityStates2021} such as higher surface resistance and lower $H_{sh}$, which can potentially cause localized heating or premature cavity quench. The location of the defects in the film largely determines their impact on cavity performance. If the defects are located near the surface where the magnetic field is strong, their impact will be far greater than if they are located deep beneath the surface where the magnetic field is close to zero due to supercurrent screening.

Sn deficient regions are theorized to form during the nucleation phase of the Vapor Diffusion process. During this phase the $Nb_3Sn$ nucleation sites are very small leading to rapid diffusion of Sn out of the $Nb_3Sn$ phase and into the Nb substrate which exceeds the rate of Sn vapor deposition. The regions have a Sn concentration of approximately 17 percent which is a stable Nb-Sn phase at low Sn concentrations. These regions have a size of around 200~\si{\nano\meter} and have been shown to reduce the surface superconducting gap and $T_c$ using point contact tunneling measurements.\cite{beckerAnalysisNb3SnSurface2015}

\section{Experimental}

\subsection{Focused Ion Beam Tomography}

The word tomography stems from the Greek words \emph{tomos}, meaning slice or section, and \emph{graphō}, meaning to display or to describe. In the case of focused ion beam (FIB) tomography slices are created using ion beam milling to remove material from the sample, and then cross section images are acquired using scanning electron microscopy (SEM). Using Ga ion beam milling we can achieve high dimensional accuracy during sample preparation and cross-section cutting allowing for slice thickness as thin as 10~\si{\nano\meter}. The material removal rate of $Ga^+$ FIB milling is high enough to analyze sample volumes on the order of $10^3$~\si{\micro\meter\cubed}. This makes $Ga^+$ FIB an ideal choice for analyzing $Nb_3Sn$ thin films which typically have a grain diameter of 1~\si{\micro\meter} and a thickness of 3-5~\si{\micro\meter}. Once the cross-section is exposed by milling, SEM provides a plethora of methods to image and gather data. These methods include secondary and back-scatter electron emission (SE/BSE) contrast imaging, secondary X-ray emission based methods such as electron dispersive X-ray spectroscopy (EDS) which provide information on sample elemental composition, and electron back-scatter diffraction (EBSD) which can measure the crystallographic orientation of the sample. These different methods can be precisely correlated in space, which is essential for analyzing $Nb_3Sn$ thin films. Previously, it has not been easy to correlate the distribution of Sn in the film with the grain structure over a large volume which is important to understand the film growth mechanism and impact on the RF performance.

In our measurements, we use a Thermo Fisher Helios 5 FIB/SEM with a Oxford Instruments Ultim® Max EDS detector and a Symmetry® S3 EBSD detector. The sample volume preparation was performed manually and then the measurement and slicing procedure was automated using the Thermo Fisher AS\&V4 software. The data acquisition was performed using the Oxford Instruments AZtec software. The AS\&V4 software automatically interfaces with the AZtec software to trigger EDS and EBSD acquisition for each slice of the sample. 

We use a cantilever sample preparation technique for our measurements. This means that a region of interest is prepared on the edge of the sample with the volume overhanging the edge. Material is removed on the sides and underneath the volume creating a channel for the diffracted electron beam to escape and hit the EBSD detector. This is necessary so that the sample does not block the signal to the EBSD detector. During EBSD the electron beam strikes the sample at a 70\si{\degree} incident angle creating a Kikuchi diffraction pattern which can be imaged by the EBSD detector to determine the crystal phase and orientation. Secondary x-rays are simultaneously measured using the EDS detector and spectrum analyzer to measure the chemical composition.

\subsection{Sample Preparation}

The sample used in this study is a cut out of a $Nb_3Sn$ coated cavity. The cavity was coated using the vapor-diffusion method.\cite{beckerAnalysisNb3SnSurface2015, leeAtomicscaleAnalysesNb3Sn2018, pudasainiGrowthNb3SnCoating2019, pudasainiNb3SnMulticellCavity2018, pudasainiRecentResultsNb3Sn2019} Before coating, the cavity was electropolished and anodized. The coating was performed at 1200~\si{\degreeCelsius} using a pure Sn source at 1300~\si{\degreeCelsius}. SnCl is used during the nucleation phase of the coating to improve nucleation density. 

After the coating, the cavity was polished using centrifugal barrel polishing for 4 hours and then recoated with a light coating step. More information about the cavity preparation is available in the author's previous publication. \cite{viklundImprovingNb3Sn2024} Several disk shaped samples were cut from the cavity around the equator region. 

To provide a flat edge for EBSD measurement, a small section of the disk is cut off using a diamond saw to create a flat, perpendicular edge. This edge is required for EBSD, since the electron beam needs to scatter at a low incidence angle into the detector. The surface finish left by the diamond saw is still too rough for FIB tomography and the mechanical force of the saw damages the $Nb_3Sn$ film near the cut edge. This requires further ion beam milling to create a sharp and undamaged edge. 

Once the edge has been prepared, a protective Pt cap of approximately 10x10~\si{\micro\meter\squared} is deposited using ion beam deposition to protect the region of interest from ion beam damage during the rest of the preparation steps. The rest of the process is to create a cantilever structure. Two trenches are cut on either side of the region of interest to a depth greater than the desired final thickness of the cantilever. The sample is then rotated so that the flat side of the sample is perpendicular to the ion beam. An undercut is then performed under the region of interest creating a cantilever with the desired thickness. In our preparation, we chose a thickness of approximately 20~\si{\micro\meter} since the $Nb_3Sn$ film is only 3~\si{\micro\meter} thick.

After preparing the sample, two fiducial marks are created on the sample, one on the top surface near the Pt cap and one on the side of the sample. The fiducial marks are created by first depositing a thin, square platinum layer and then milling a high contrast symbol onto the layer. These marks are used by the AS\&V software to locate the sample after every stage move.

\subsection{Grain Boundary Reconstruction From Point Cloud Data}

To reconstruct the grain boundaries from our EBSD data, we use a voronoi decomposition based algorithm. This algorithm was first used by the MTEX Matlab package created by Bachmann, et. al~\cite{bachmannGrainDetection2d2011}. We have re-implemented the algorithm in the Python programming language using the Scipy and Numpy libraries. We also visualize the reconstruction using the PyVista library. This implementation of the voronoi grain reconstruction algorithm is available on GitHub\cite{viklundEBSDVoroPy2025}. 

\begin{figure*}
	\centering
	\includegraphics[width=1\linewidth]{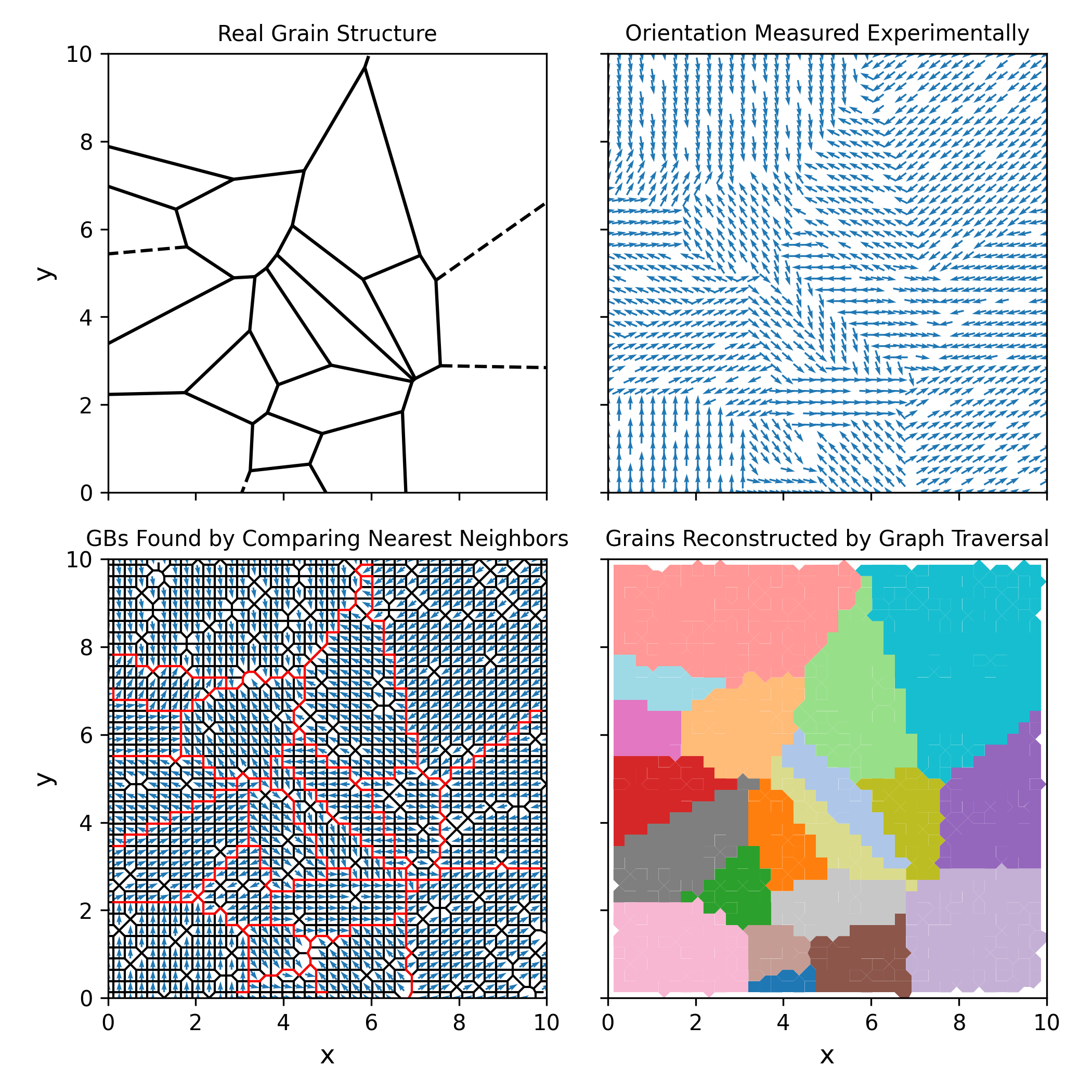}
	\caption{A diagram showing the voronoi-based reconstruction algorithm. First the grain orientation of a real sample is measured at discrete grid-like positions. Then the voronoi decomposition is calculated from the measurement positions to produce a nearest neighbor graph. The misorientation angle is calculated between nearest neighbor cells and GBs are determined using a minimum threshold angle. The grains are then reconstructed by traversing the graph and finding all regions that are enclosed by a GB.}
	\label{fig:algodiagram}
\end{figure*}

As shown in figure \ref{fig:algodiagram}, the algorithm constructs a voronoi diagram to sort the data into a nearest neighbor graph based on the position of the data points. The voronoi diagram separates the space into regions, Voronoi cells, that contain all points in space that are nearest to the cell's corresponding data point. Figure \ref{fig:algodiagram} only shows a 2D example, but the actual algorithm operates on 3D data points which means the boundaries of the voronoi cells are actually 2D polygons and the cells are 3D convex polygonal volumes. By connecting the points whose voronoi cells share a face we can construct a nearest neighbor graph, which is the mathematical dual of the voronoi diagram. This graph can also be called the Delaunay triangulation of the data points where the edges of the triangulation correspond to the edges of the graph. 

To find the GBs in our sample, we calculate the misorientation angle associated with each edge in the Delaunay triangulation, that is, the angle between the grain orientations of every pair of voronoi cells that share a face. If the misorientation angle exceeds a threshold value then we label that shared edge a GB or a sub-GB. The grain can then be found by traversing the nearest neighbor graph without crossing any GBs/sub-GBs. A single grain is found by the union of all voronoi cells that can be reached from a single randomly selected starting point without crossing a GB. This process is repeated for any cell that has not yet been associated with a grain until all regions have been associated with a grain. The distinction between a GB and a sub-GB is that GBs define the border between two separate grains while both sides of a sub-GB are the same grain, i.e. there is a path around the sub-GB that does not cross any other GBs.

This algorithm, while more complex to implement, greatly simplifies the analysis, since data points do not need to adhere to a structured grid like traditional methods. This makes our analysis robust against unidentified data points in the EBSD data, which can simply be removed. This algorithm can also be applied to points that do not adhere to a simple grid which can happen if the sample slicing plane is not perpendicular to the surface. Additionally, some EBSD imaging software may allow for hexagonal grid imaging to reduce grid artifacts. It can even work for completely unstructured point cloud data such as that produced by atom probe tomography (APT). Since the resulting grain boundaries are actual mesh structures rather than a voxel grid, we can apply mesh based calculations like finding the distance of a point from the GB easily. We also have the ability to slice the mesh along arbitrary planes to analyze the cross section which would be much more difficult with a voxel mesh.

\section{Results}

\begin{figure*}
	\centering
	\includegraphics[width=0.9\linewidth]{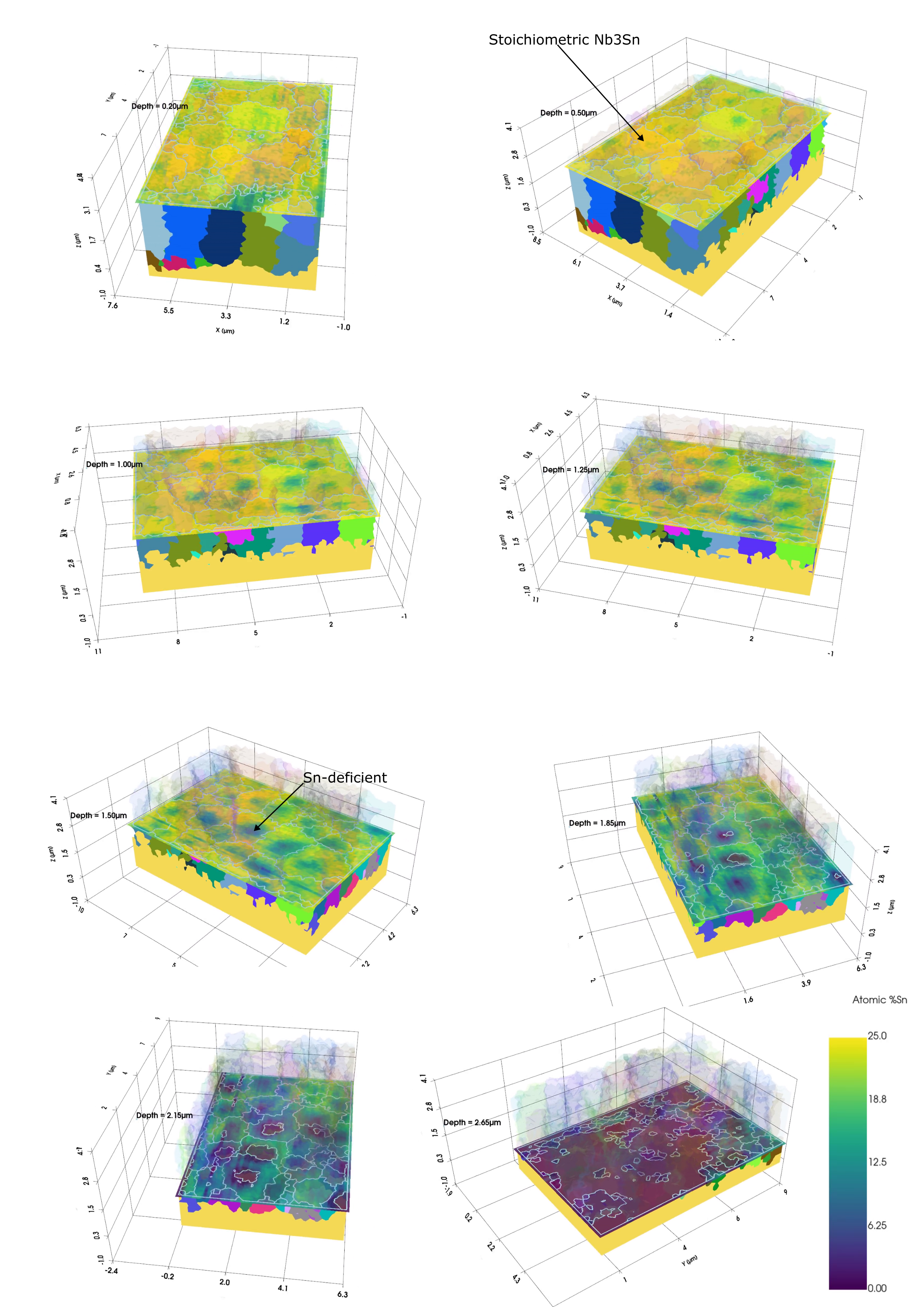}
	\caption{A 3D reconstruction of the Nb3Sn grains. Each grain is colored with a unique color. A horizontal plane is used to slice the sample. The plane is colored according to the Sn concentration at each point on the plane surface. }
	\label{fig:horizontal_slices}
\end{figure*}

Our analysis shows that the Sn deficient regions are present below the surface in nearly every grain that we analyzed. The regions are located in the centers of the grains while the grain boundaries are slightly Sn rich. In figure~\ref{fig:horizontal_slices} you can see the Sn distribution in the film in relation to the grain boundaries at various depths below the surface. 

\begin{figure}
	\centering
	\includegraphics[width=1.0\linewidth]{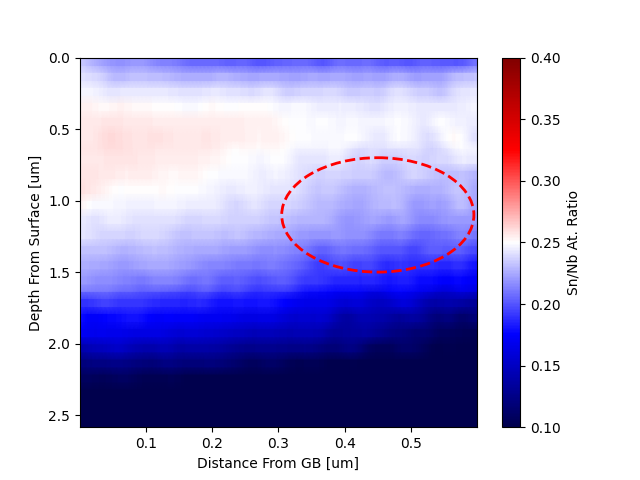}
	\caption{In this figure we show the Sn concentration in a $Nb_3Sn$ film as a function of depth and distance from nearest grain boundary. Blue areas indicate Sn concentration below stoichiometric $Nb_3Sn$ and red areas above. The areas of highest Sn concentration are found most commonly near grain boundaries in the top half of the film and areas of low Sn concentration are found far from GBs below 500~\si{\nano\meter} from the surface. These Sn deficient regions are indicated by a red ellipse. The top 100~\si{\nano\meter} of the film also appear to have low Sn content however this is likely caused by the poor spatial resolution of the EDS measurement.}
	\label{fig:depthanalysis}
\end{figure}

As shown in figure~\ref{fig:depthanalysis} the Sn deficiency becomes more severe the farther from the top surface you go. Below a depth of around 1.5~\si{\micro\meter} from the surface all areas are Sn deficient due to the proximity of the Nb Substrate layer. We find that the Sn deficiency occurs around 1.0~\si{\micro\meter} below the surface far away from the GBs near the center of the grains, but at a depth above 500~\si{\nano\meter} the film is not Sn deficient. The figure appears to show that the film is also Sn deficient on the surface, however this is due to the size of the electron beam interaction volume which causes a blurring effect with the vacuum above the film leading to a lower Sn signal. 

Closer to the GB the film may appear slightly Sn rich. This is not due to Sn segregation at the GBs, which has been observed in previous studies~\cite{leeGrainboundaryStructureSegregation2020}. This effect occurs on a length scale of 3~\si{\nano\meter} which is far too small to resolve using our measurement technique. The measured Sn concentration should also not be considered completely accurate, since we have not calibrated the measured Sn concentration to a known standard.

\begin{figure*}
	\centering
	\includegraphics[width=1.0\linewidth]{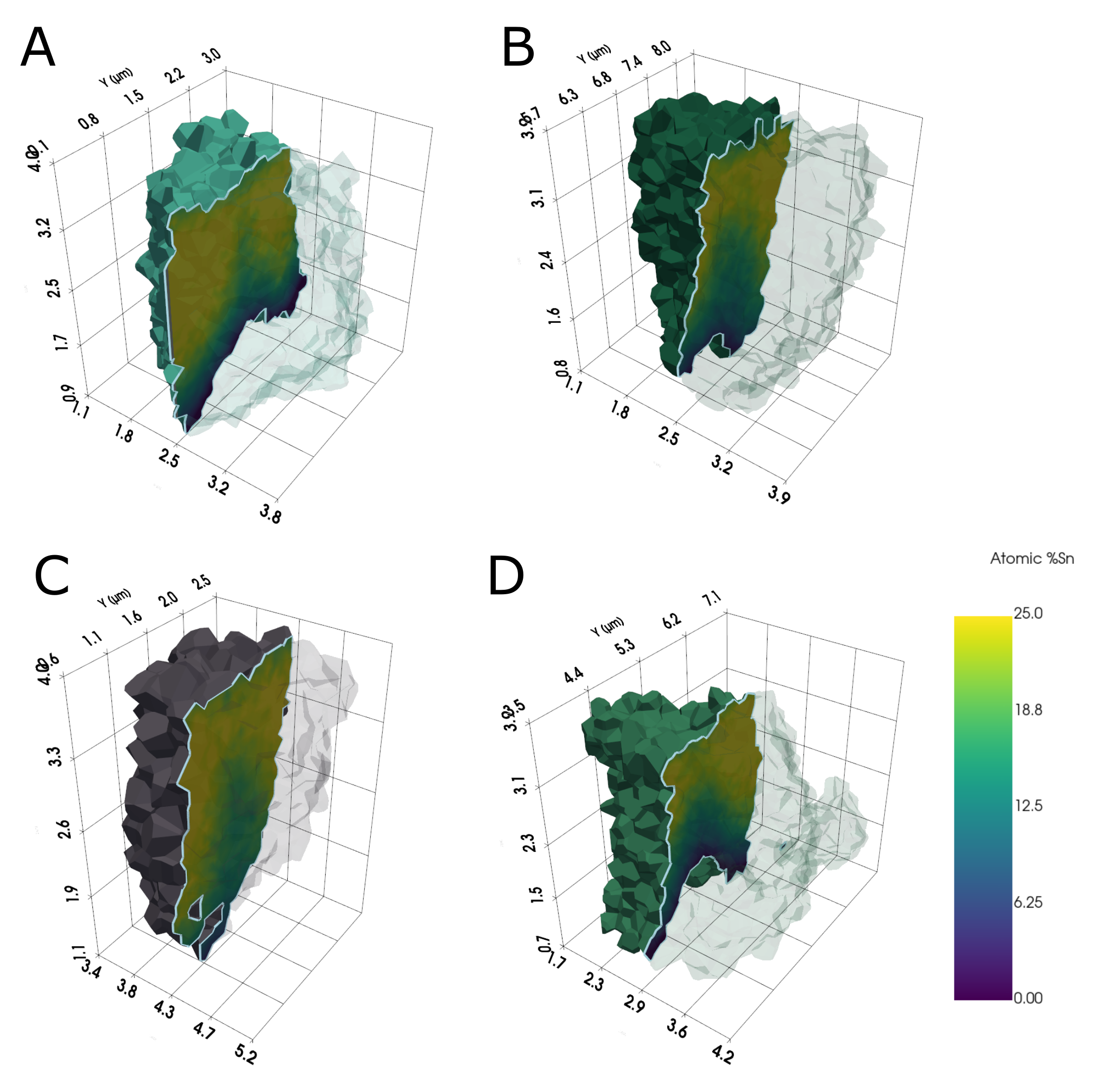}
	\caption{Cross-sectional view of four different $Nb_3Sn$ grains. The cross section area is colored based on the Sn content. Grains consistently show Sn deficiency near the Nb substrate and the center of the grain far from GBs.}
	\label{fig:grainsections}
\end{figure*}

\begin{figure*}
	\centering
	\includegraphics[width=1.0\linewidth]{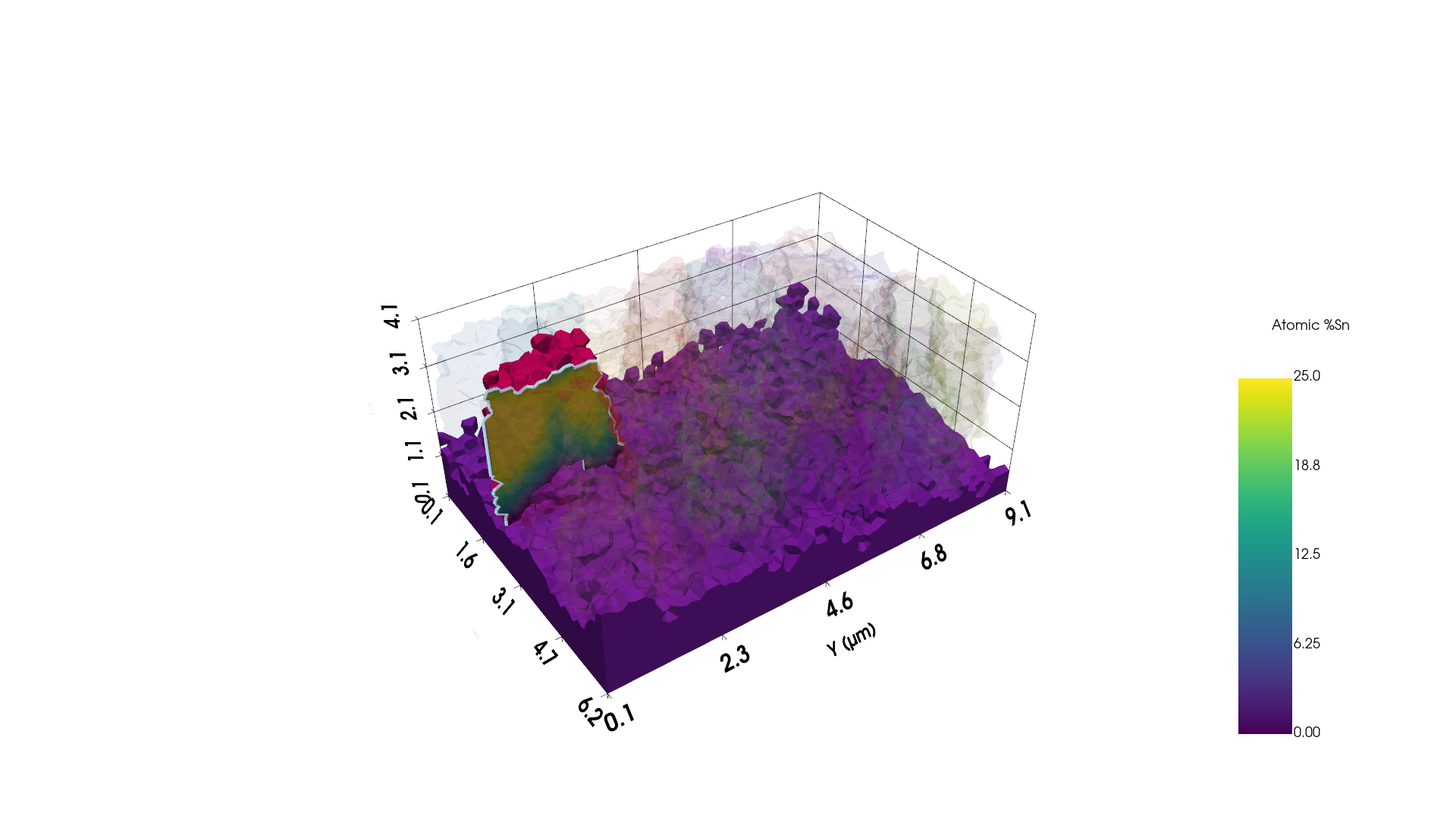}
	\caption{A cross-sectional view of a $Nb_3Sn$ grain and the Nb substrate. The cross section is colored based on Sn concentration. The rest of the $Nb_3Sn$ grains are rendered translucent for reference.}
	\label{fig:grainsectionwcontext24}
\end{figure*}

In figures~\ref{fig:grainsections} and \ref{fig:grainsectionwcontext24} we show cross sections of individual $Nb_3Sn$ grains sliced vertically along the Y-Z plane through their center of mass. This figure shows that the Sn deficiency is consistently found in the centers of the grain. The deficiency is greatest near the Nb substrate but also extends closer to the surface away from the GBs. We also see that the bottom surface of the grains are scalloped, which is consistent with our expectations based on the GB short circuit diffusion growth model\cite{pudasainiGrowthNb3SnCoating2019}.

\section{Discussion}

The presence of Sn deficient regions has been known for some time\cite{beckerAnalysisNb3SnSurface2015,leeAtomicscaleAnalysesNb3Sn2018,trenikhinaPerformancedefiningPropertiesNb3Sn2017} from 2D cross-section analysis. However, these regions were assumed to be rare. Using 3D measurements it is now possible to see how these regions penetrate into different parts of the grain, which was impossible in 2D. This has clearly showed that Sn deficient regions are found more frequently than not in the $Nb_3Sn$ film. We can also see how these regions lie relative to GBs and interfaces. We find that they exist most commonly in the centers of grains rather than near GBs. Furthermore, we see that these regions extend close to the film surface within 1~\si{\micro\meter} rather than only occurring near the Nb interface.

\subsection{Origin of Sn Deficient Regions}

Sn deficient regions have previously been theorized to be caused by the remnants of the initial $Nb_3Sn$ nucleation sites\cite{leeAtomicscaleAnalysesNb3Sn2018}. However, due to the structure of these regions they may in fact be caused by the growth process instead. $Nb_3Sn$ film growth is known to occur via GB short circuit diffusion\cite{pudasainiGrowthNb3SnCoating2019} which means the diffusion rate of Sn is much higher through the GBs compared with through the bulk $Nb_3Sn$. This mechanism leads to a cusped interface between the film and the substrate, a phenomenon that we also observe in our measurements. Additionally, the Sn concentration near the substrate interface is always lower than 25\% due to the formation of the stable 17\% Nb-Sn phase. This Sn deficient phase is enriched by Sn from the GBs, but regions that are not close to a GB remain Sn deficient. This leads to every grain containing a Sn deficient core that extends up close to the surface. Near the surface, Sn is supplied by bulk diffusion from the surface of the film exposed to Sn vapor. Our measurements suggest that these regions extend to within 1~\si{\micro\meter} of the surface, however it is hard to know the exact extent due to the poor spatial resolution of the EDS measurement. More accurate 2D TEM EDS measurements of these regions indicate that they are localized to the substrate interface\cite{beckerAnalysisNb3SnSurface2015}, however these measurements cannot probe a large statistically significant sample volume like we have demonstrated.

\subsection{Effects on $Nb_3Sn$ Cavity Performance}

It is unlikely that changes to the Sn vapor diffusion coating process can eliminate Sn deficient regions, since they are a product of the growth mechanism itself. However, the impact of these regions on the RF performance of the coating can be mitigated. The magnitude of the electromagnetic fields inside a superconductor decay exponentially proportional to $\exp^{-\frac{z}{\lambda}}$ where $z$ is the distance from the surface and $\lambda$ is the London penetration depth of the material. Since the penetration depth of $Nb_3Sn$ is on the order of 100\si{\nano\meter}, the field will be strongly attenuated bellow 1-200~\si{\nano\meter}. As long as the top most layer of the film does not contain Sn deficient regions they will not interact strongly with the RF field. 

\subsection{Implications for $Nb_3Sn$ Cavity Surface Polishing}

Surface polishing techniques such as centrifugal barrel polishing (CBP) have recently shown promising results for improving $Nb_3Sn$ cavity performance\cite{viklundImprovingNb3Sn2024}. However, these performance improvements necessitate a post polishing Sn recoating step to eliminate the severe gradient and quality factor degradation caused by the polishing initially. The cause of this initial degradation is unknown. It may be caused by fractures in the $Nb_3Sn$ film due to stress applied to the cavity during CBP. The recoating step has been shown to heal cracks in $Nb_3Sn$ films caused by plastic deformation of the Nb substrate\cite{viklundHealingGradientDegradation2024}, which could explain the initial degradation and recovery. Our current study suggests another possible cause for performance degradation after polishing. The material removal during polishing may remove the stoichiometric $Nb_3Sn$ top layer of the film and expose Sn deficient regions below the surface of the film. Once exposed, these regions interact with the RF field and reduce the superheating field and increase the surface resistance of the film. By recoating the cavity, these exposed Sn deficient regions can absorb Sn and transform into stoichiometric $Nb_3Sn$.

\subsection{Recommendations for Future Research}

Our results show that the properties of Sn deficient regions in $Nb_3Sn$ thin films and their formation mechanism is still poorly understood. To improve our understanding of this phenomenon, we recommend further research into the effects of different coating parameters on these regions. For example, to create a thicker stoichiometric top layer it would be beneficial to increase the coating duration allowing Sn to diffuse deeper into the $Nb_3Sn$ grains, however this has other negative effects such as increasing surface roughness and total film thickness. Another option could be to increase the coating temperature which would increase the diffusion rate of Sn through the bulk $Nb_3Sn$. In general it seems intuitive that a film with larger grains would have fewer Sn deficient regions since as the GBs move horizontally through the film they should convert any Sn deficient regions into stoichiometric material by allowing Sn to diffuse more quickly. Simultaneously, the remaining Sn deficient regions are expected to shrink over time due to slow diffusion of Sn through the bulk from the surface and nearby GBs. However, this cannot be proven without further measurements on different samples or simulation of the grain coarsening and short circuit diffusion system.

\section{Conclusion}

In this study we have used FIB Tomography to analyze Sn vapor diffusion coated $Nb_3Sn$ films in 3D with crystallographic and chemical measurements. We use a voronoi diagram based algorithm to reconstruct the grain structure of the sample volume implemented in a custom Python program. This is the first large volume analysis of both chemical composition and grain structure of $Nb_3Sn$ thin film in 3D, which we use to correlate the distribution of Sn deficient regions in the film with the location of grain boundaries. We find that Sn deficient regions are much more prevalent in the film than previously thought. We find that nearly every grain in our sample volume contains a Sn deficient region near the center of the grain that extends down to the Nb substrate. The Sn deficiency is estimated to start between 0.5~\si{\micro\meter} to 1.0~\si{\micro\meter} from the surface. Due to the much shorter penetration depth of $Nb_3Sn$, approximately 100~\si{\nano\meter}, these Sn deficient regions should not have a major impact on the performance of $Nb_3Sn$ cavities. However, if the top surface of the film is removed by polishing a recoating is necessary to restore the Sn concentration in the RF layer. This would explain why we initially observe a performance degradation in polished $Nb_3Sn$ until a recoating is performed.\cite{viklundImprovingNb3Sn2024}

\section{Acknowledgements}

This manuscript has been authored by FermiForward Discovery Group, LLC under Contract No. 89243024CSC000002 with the U.S. Department of Energy, Office of Science, Office of High Energy Physics.

\printbibliography

\end{document}